\documentclass[a4paper,12pt]{article}
\usepackage[dvips]{graphicx}
\usepackage{amsmath}
\begin{document}
\begin{center}
{\large \bf The fractal cosmological model}
\vspace{5ex}

I. K. Rozgacheva$^a$, A. A. Agapov$^b$
\vspace{2ex}

Moscow State Pedagogical University, Moscow, Russia
\vspace{1ex}

E-mail: $^a$rozgacheva@yandex.ru, $^b$agapov.87@mail.ru
\vspace{4ex}
\end{center}
\begin{abstract}

The fractal cosmological model which accounts for observable fractal properties of the Universe's large-scale structure is constructed. In this framework these properties are consequences of the rotary symmetry of charged scalar meson matter field (complex field). They may be explained through a conception of the Universe as an assembly of self-similar space-time domains. We have found the scale invariant solutions of Einstein's equation and Lagrange's field equation. For the solution the space-time domains with field values related by the scaling are geometrically similar and evolve similarly. Due to this, fractal properties of initial density perturbations which lead to formation of the large-scale structure remain during the gravitational evolution and lead to presence of the fractal properties in the large-scale structure.
\end{abstract}
\vspace{1ex}

KEY WORDS: complex field, rotary symmetry, fractal properties of the large-scale structure, fractal cosmological model
\vspace{3ex}

\begin{center}
{\bf 1. Introduction}
\end{center}

The analysis of available at present galaxy catalogues shows that the galaxy distribution possesses a range of fractal properties \cite{1}, \cite{2}, \cite{3}, \cite{4}, \cite{5}, \cite{6}, \cite{7}.

- The observable angular correlation function of the galaxy distribution on the celestial sphere is a power law
$$
\omega(\vartheta)\sim\vartheta^{-\gamma},\eqno (1)
$$
where $0{,}6\le\gamma\le1{,}2$ depending on a catalogue, $\vartheta$ is angular distance between galaxies.

- In clumps the dependence of number of galaxies within distance less than r on the distance r is a power law as well
$$
N\left(\le r\right)\sim r^{d_c},\eqno (2)
$$
where the correlation dimension $d_c$ takes values $d_c\approx 1{,}15\div 2{,}25$ at scales $r\approx \left(1\div 10\right)h^{-1} Mpc$.

In the previous paper \cite{8} the analysis of the quasar distribution according to the seventh version of the largest at present survey SDSS has been carried out. This distribution possesses fractal properties as well.

- Within the redshift range $0{,}35\le z\le 2{,}30$ the dependence of number of quasars in a volume with radius $r$ centered at the observer on $r$ is found to be a power law (2) with $d_c=2{,}71$ for the flat Universe filled with cold dust.

- For each quasar located approximately at the center of a concentration region on the celestial sphere number of neighbour quasars located within angular distances less than $\vartheta$ is possessed of a power-law dependence on $\sin{\vartheta/2}$:
$$
N\left(\le \vartheta\right)\sim\left(\sin{\frac{\vartheta}{2}}\right)^{d_c},\eqno (3)
$$
where $d_c\approx 1{,}49\div 1{,}58$ for different redshift layers in the same redshift range.

The aim of this work is construction of the cosmological model permitting physical explanation of fractal properties (1) - (3) of the large-scale structure of the Universe.

What these properties of the large-scale structure are indicative of? The obvious answer is that the properties are consequences of the fractal properties of the initial matter density perturbations which further led to star, galaxy and cluster formation due to gravitational instability. Now we observe traces of these fractal properties through quasars.

This interpretation follows from the hypothesis of Gaussian (thermal) spectrum of the initial density perturbations. Let's consider a spherical volume V containing mass M in continuous medium. Probability of appearance of thermal density fluctuation near this mass is defined  by the formula \cite{9}, \cite{10}
$$
W\sim \exp\left[-\frac{c_v}{2}{\left(\frac{\delta T}{T}\right)}^2-\frac{M}{2kT}{\left(\frac{\partial P}{\partial \rho}\right)}_T{\left(\frac{\delta V}{V}\right)}^2-\frac{R_{min}}{kT}\right],
$$
where $\rho$ is medium density, $P$ is pressure, $c_v$ is medium heat capacity at constant volume, $\delta T$ and $\delta V$ are independent fluctuations of temperature and volume respectively, $R_{min}$ is minimal work necessary for reversible removal of mass $\delta M$ for distance $\delta r$ in the gravity field of mass $M$. In case of radial displacement the work equals
$$
R_{min}\approx G\frac{M\cdot \delta M}{r^2}\delta r=\frac{4\pi}3G\rho r\cdot \delta M\cdot \delta r
$$
in Newtonian approximation, where $\delta M = 4\pi \rho r^2\delta r$ is a mass of a spherical layer. In this case, probability of a thermal fluctuation equals
$$
W=\frac1{2\pi \Delta_T\Delta_r}\exp{\left[-\frac1{2{\Delta_T}^2}{\left(\frac{\delta T}{T}\right)}^2-\frac1{2{\Delta_r}^2}{\left(\frac{\delta r}{r}\right)}^2\right]},
$$
where the variances equal
$${\Delta_T}^2={c_v}^{-1},\ \ \ {\Delta_r}^2={\left\{\frac{12\pi\rho}{kT}\left[{\left(\frac{\partial P}{\partial\rho}\right)}_Tr^3+\frac{8\pi G\rho}9r^5\right]\right\}}^{-1}.
$$

Fluctuation of density. in a volume $V$ equals
$$
\delta\rho=\delta\left(\frac{M}{\frac{4\pi}{3}r^3}\right)=-3\rho\left(\frac{\delta r}r\right).
$$
The root-mean-square relative density fluctuation (fluctuation spectrum) equals
$$
\sqrt{\left\langle\left(\frac{\delta\rho}{\rho}\right)^2\right\rangle}=3\sqrt{\left\langle\left(\frac{\delta r}r\right)^2\right\rangle}=3{\left\{\frac{12\pi\rho}{kT}\left[{\left(\frac{\partial P}{\partial\rho}\right)}_Tr^3+\frac{8\pi G\rho}9r^5\right]\right\}}^{-1/2}.\eqno (4)
$$
At spatial scales for which pressure gradients are important, i.e. when the first term in the brackets dominates, one have the "white noise" spectrum (Zel'dovich-Harrison spectrum):
$$
\sqrt{\left\langle\left(\frac{\delta\rho}{\rho}\right)^2\right\rangle}\sim r^{-1{,}5}.
$$
At large scales for which gravity effects are important, i.e. when the second term dominates, one has the following spectrum
$$
\sqrt{\left\langle\left(\frac{\delta\rho}{\rho}\right)^2\right\rangle}\sim r^{-2{,}5}.
$$
The spectrum (4) is scale invariant because the fraction $\frac{\delta\rho}{\rho}$ doesn't change under scale transformations. This thermal fluctuations spectrum is an example of fractal spectrum.

The root-mean-square relative density fluctuation is an estimate of the correlation function according to the correlation function definition \cite{2}:
$$
\xi =\left\langle\left(\frac{\delta\rho}{\rho}\right)^2\right\rangle^{1/2}.
$$
For a random fluctuation average number of neighbour fluctuations within distance less then r may be estimated as
$$
\left\langle N\right\rangle\approx 4\pi \left\langle n\right\rangle \int\limits_0^r\left(1+\xi\right){\tilde r}^2d\tilde r,
$$
where $n$ is mean fluctuations number density. In case of fluctuations clumping and $\xi\ge1$ there are fractal laws like (2): $\left\langle N\right\rangle \sim r^{1{,}5}$ for the white noise and $\left\langle N\right\rangle \sim r^{0{,}5}$ at large scales.

The observable correlation dimension value $d_c\approx 1{,}15\div 2{,}71$ may follow from the spectrum
$$
\sqrt{\left\langle\left(\frac{\delta\rho}{\rho}\right)^2\right\rangle}\sim r^{-1{,}85}\div r^{-0{,}29},
$$
therefore, it permits of not only the white noise spectrum.

Thereby, the fractal laws (1) - (3) are expected to exist in Newtonian approximation. However, it is not quite so in the general theory of relativity because Einstein's tensor is not invariant under scale transformation of the Riemannian space-time \cite{11}, \cite{12}. If the large-scale structure evolution is described by Einstein's gravity theory the fractal properties may not conserve, even if the initial fluctuations had the thermal spectrum.

The gravity theory in Riemannian spaces can become scale invariant if an additional non-Newtonian interaction described by a scalar field (dilaton) is introduced. In this case field equations are invariant under transformations of the Weyl group (conformal transformations), i. e. the local interval transformation $ds^2\to \sigma(x)ds^2$ and the local field transformation $\Phi \to \sigma^{1/2}\Phi$. The non-Newtonian interaction is different in gravitational systems with different scales and one can attain presence of the fractal properties through this interaction. But no evidences of existing dilaton field have been revealed.

In gauge gravity theories the scale invariance may be achieved through conversion to post-Riemannian spaces and introduction of additional matter interactions in fact (see the review of such type theories in book \cite{13}). The question of existence of these interactions is open.

In this paper the hypothesis that the observational fractal properties of the large-scale structure may be a consequence of existence of the matter fields rotary symmetry is advanced. The charged scalar meson field (complex field) with rotary symmetry
$$
\psi\psi^*=\Psi^2=\mbox{const},\eqno (5)
$$
(where the asterisk denotes complex conjugation and $\Psi$ is field amplitude relating to the field charge $Q\sim\Psi^2$) is an opportune example. In this case Einstein's and Lagrange's equations turn out to be satisfied for the class of fields $\psi$ and $\tilde\psi$ which possess constant energy densities and their phases are related by the transformation:
$$
\varphi\leftrightarrow\tilde\varphi+\alpha,\eqno (6)
$$
where $\alpha$ is a numerical transformation parameter.

In this case field energy densities and space-time metric tensors differ in a constant factor only:
$$
E\leftrightarrow\frac{U_0}{\tilde U_0}\tilde E,
$$
$$
g_{mn}\left(\psi\right)\leftrightarrow\frac{\tilde U_0}{U_0}\tilde g_{mn}\left(\tilde \psi\right),
$$
where $U_0$ and $\tilde U_0$ are the constant field potential parameters, $U=U_0\psi\psi^*$, $\tilde U=\tilde U_0\tilde \psi\tilde \psi^*$. Therefore, space-time volumes with field values related by the scaling (6) are geometrically similar and evolve similarly. The symmetry (5) is conserved under the scaling (6).

The power law (2) arises for fields if their energy densities are related by a power transformation:
$$
E\leftrightarrow\left(\tilde E\right)^{\beta}, \eqno (7)
$$
where $\beta$ is a numerical transformation parameter. Indeed, a measure of field condition is proportional to field energy $E$, on the one hand. On the another hand, this measure is proportional to a ratio of space volume $V$ occupied by the field to whole observable volume $r^3$. Therefore $E\sim \frac{V}{r^3}$. Let there are $N$ volumes $\tilde V$ occupied by the field with energy density $\tilde E$ and $\frac{\tilde V}{r^3}\approx \frac1{N}$, $\tilde V\sim V^{1/\beta}$. Let interaction between volumes may be neglected, then:
$$
\tilde E\sim E^{1/\beta}\sim \left(\frac{V}{r^3}\right)^{1/\beta}\sim \left(\frac{\tilde V}{r^3}\right)r^{3-3/\beta}\sim \frac1Nr^{3-3/\beta}.
$$
Hence, a power law like (2) for volumes with field amplitudes $\tilde\Psi$ is obtained:
$$
N\left(\le r\right)\sim \frac1{\tilde E}r^{3-3/\beta}.
$$
\vspace{1ex}

\begin{center}
{\bf 2. The solution of Einstein's and Lagrange's equations for complex field}
\end{center}

Let's consider a dynamic system of gravity and complex $\psi$ fields described by Einstein-Hilbert action within general relativity framework:
$$
S=-\frac{c^3}{16\pi G}\int\left(R-\frac{8\pi G}{c^4}L\right)\sqrt{-g}\,d^4x,
$$
where $R$ is scalar curvature, $g<0$ is determinant of metric  tensor $g_{mn}$, space-time interval is $ds^2=g_{mn}dx^mdx^n$, indices take values 0, 1, 2, 3, metric signature is $(+---)$. We use the following form of complex field Lagrangian:
$$
L=\frac1{hc}\left(g^{mn}\frac{\partial\psi}{\partial x^m}\frac{\partial\psi^*}{\partial x^n}-U\left(\psi\psi^*\right)\right),\eqno (8)
$$
where $U\left(\psi\right)$ is field potential, $h$ is Planck's constant, $c$ is light velocity. Hereafter, the field dimension is $\left[\psi\right]=\mbox{erg}$, the contravariant metric tensor dimension is $\left[g^{mn}\right]=\mbox{cm}^{-2}$. This field possess the symmetry (5). Its Lagrange equation is
$$
\frac1{\sqrt{-g}}\frac{\partial}{\partial x^n}\left(\sqrt{-g}g^{mn}\frac{\partial\psi}{\partial x^m}\right)=-\frac{\partial U}{\partial\psi^*}.\eqno (9)
$$

In Einstein's equation
$$
R_n^m-\frac12R\delta_n^m=\kappa T_n^m\eqno(10)
$$
energy-momentum tensor of the complex field equals
$$
T_n^m=\frac{\partial\psi}{\partial x^n}\frac{\partial L}{\partial \left(\frac{\partial\psi}{\partial x^m}\right)}+\frac{\partial\psi^*}{\partial x^n}\frac{\partial L}{\partial \left(\frac{\partial\psi^*}{\partial x^m}\right)}-\delta_n^mL=\frac1{hc}g^{mp}\left(\frac{\partial\psi}{\partial x^p}\frac{\partial\psi^*}{\partial x^n}+\frac{\partial\psi}{\partial x^n}\frac{\partial\psi^*}{\partial x^p}\right)-\delta_n^mL,
$$
where $R_n^m$ is Ricci tensor, $\kappa =\frac{8\pi G}{c^4}$ is Einstein's gravity constant, $G$ is Newton gravity constant, $\delta_n^m$ is delta symbol.

Following form of potential is used further:
$$
U=U_0\psi\psi^*.\eqno (11)
$$
One can ascertain that Lagrange's equation (9) with potential (11) is satisfied for the solution:
$$
\psi =\Psi\mbox{e}^{i\varphi},\ \ \ \psi^*=\Psi\mbox{e}^{-i\varphi},
$$
$$
\Gamma_{mn}^l=\frac1{U_0}\frac{\partial^2\varphi}{\partial x^m\partial x^n}\left(g^{lp}\frac{\partial\varphi}{\partial x^p}+a^l\right),\eqno (12)
$$
$$
g_{mn}=\frac1{U_0}\left(4\frac{\partial\varphi}{\partial x^m}\frac{\partial\varphi}{\partial x^n}+\frac{\partial\varphi}{\partial x^m}a_n+\frac{\partial\varphi}{\partial x^n}a_m\right),
$$
where field phase $\varphi\left(x^m\right)$ is a differentiable function. Hereafter, indices are raised and lowered with the metric tensor, indices appearing twice in a single term imply summing over its values, semicolon denotes covariant differentiation, $\Gamma_{mn}^l$ are Christoffel symbols.

Derivative $\displaystyle \frac{\partial\varphi}{\partial x^m}$ and covariant vector $a_m$ satisfy equations:
$$
g^{mn}\frac{\partial\varphi}{\partial x^m}\frac{\partial\varphi}{\partial x^n}=U_0,
$$
$$
g^{mn}\left(\frac{\partial\varphi}{\partial x^m}\right)_{;n}=0,
$$
$$
a_{m;l}=0,\ \ \ a_ma^m=-3U_0,\eqno (13)
$$
$$
\frac{\partial\varphi}{\partial x^m}a^m=0.
$$
Covariant $a_m$ and contravariant $a^k$ vectors satisfy equations:
$$
\frac{\partial a_m}{\partial x^l}=-3\frac{\partial^2\varphi}{\partial x^m\partial x^l},
$$
$$
\frac{\partial a^n}{\partial x^m}a_n=3a^n\frac{\partial^2\varphi}{\partial x^n\partial x^m}.\eqno (14)
$$
One can ascertain through a substitution that the following equalities are satisfied for the solution (12 - 14):
$$
\frac{\partial g_{mn}}{\partial x^l}=g_{km}\Gamma_{nl}^k+g_{kn}\Gamma_{ml}^k,
$$
$$
\Gamma_{kl}^m=\frac12g^{mn}\left(\frac{\partial g_{nk}}{\partial x^l}+\frac{\partial g_{nl}}{\partial x^k}-\frac{\partial g_{kl}}{\partial x^n}\right),
$$
$$
\delta_m^n=g^{nl}g_{lm}.
$$
The Ricci tensor and the energy-momentum tensor for this solution equals:
\begin{multline*}
R_{km}=g^{jl}R_{jklm}=\frac{\partial \Gamma_{km}^l}{x^l}-\frac{\partial \Gamma_{kl}^l}{x^m}+\Gamma_{km}^l\Gamma_{ln}^n-\Gamma_{kl}^n\Gamma_{mn}^l=\\
\shoveleft{=\frac1{U_0}\left(\frac{\partial^2\varphi}{\partial x^k\partial x^m}\frac{\partial a^l}{\partial x^l}-\frac{\partial^2\varphi}{\partial x^k\partial x^l}\frac{\partial a^l}{\partial x^m}\right)+}\\
\shoveleft{{+\frac1{{U_0}^2}\left(\frac{\partial^2\varphi}{\partial x^k\partial x^m}\frac{\partial^2\varphi}{\partial x^n\partial x^l}-\frac{\partial^2\varphi}{\partial x^k\partial x^l}\frac{\partial^2\varphi}{\partial x^n\partial x^m}\right)\left(g^{lp}\frac{\partial\varphi}{\partial x^p}+a^l\right)a^n,}}
\end{multline*}
$$
T_{km}=\frac2{hc}\Psi^2\frac{\partial\varphi}{\partial x^k}\frac{\partial\varphi}{\partial x^m}.
$$
Functions $\displaystyle \frac{\partial\varphi}{\partial x^m}$, $a_m$, $a^m$ are determined by equations (10) and (14).

The expression for Ricci tensor implies the second derivative $\frac{\partial^2\varphi}{\partial x^m\partial x^n}$ is a tensor. For example, it may be so if the function $\varphi$ depends on an argument $y=k_mx^m$, where $k_m$ is a wave vector which is parallel transferred along a geodesic line: $k_{m;n}=0$. In this case one can derive:
$$
\frac{\partial^2\varphi}{\partial x^m\partial x^n}=\frac{\frac{d^2\varphi}{dy^2}}{1-\frac1{U_0}\left(\frac{d\varphi}{dy}\right)^2\left(k^pk_p\right)}k_mk_n.
$$
This expression shows that $\displaystyle \frac{\partial^2\varphi}{\partial x^m\partial x^n}$ is proportional to a product of two vectors, therefore it's a tensor.

For example, consider the solution
$$
\frac{\partial^2\varphi}{\partial x^k\partial x^m}=b_k\frac{\partial\varphi}{\partial x^m}+b_m\frac{\partial\varphi}{\partial x^k},\eqno (15)
$$
where the vector $b_m$ satisfies the condition
$$
b_m\frac{\partial\varphi}{\partial x^m}=0.
$$
Then the following expressions for Ricci tensor and scalar curvature are obtained:
$$
R_{km}=\frac1{U_0}\frac{\partial^2\varphi}{\partial x^k\partial x^m}\left(\frac{\partial a^l}{\partial x^l}+a^nb_n\right)-\frac1{U_0}\frac{\partial\varphi}{\partial x^k}\frac{\partial a^l}{\partial x^m}b_l-\left(\frac{a^nb_n}{U_0}\right)^2\frac{\partial\varphi}{\partial x^k}\frac{\partial\varphi}{\partial x^m},
$$
$$
R=-\frac1{U_0}g^{km}\frac{\partial\varphi}{\partial x^k}\frac{\partial a^l}{\partial x^m}b_l-\frac{\left(a^nb_n\right)^2}{U_0}.
$$
Einstein's equation (10) results in the following equations:
$$
\frac1{{U_0}^2}g^{km}\frac{\partial\varphi}{\partial x^k}\frac{\partial a^l}{\partial x^m}b_l+\left(\frac{a^nb_n}{U_0}\right)^2=\frac{2\kappa}{hc}\Psi^2,\eqno (16)
$$
\begin{multline*}
\frac{\partial^2\varphi}{\partial x^k\partial x^m}\left(\frac{\partial a^l}{\partial x^l}+a^nb_n\right)-\frac{\partial\varphi}{\partial x^k}\frac{\partial a^l}{\partial x^m}b_l+\frac{\kappa}{hc}\Psi^2U_0\left(a_k\frac{\partial\varphi}{\partial x^m}+a_m\frac{\partial\varphi}{\partial x^k}\right)+\\
+\left(\frac{2\kappa}{hc}\Psi^2U_0-\frac{\left(a^nb_n\right)^2}{U_0}\right)\frac{\partial\varphi}{\partial x^k}\frac{\partial\varphi}{\partial x^m}=0.
\end{multline*}
The equations (14) and (16) determine the class of Einstein's and Lagrange's equations solutions for a complex field possessing the symmetry (5) when two parameters $\Psi$ and $U_0$ and boundary conditions are specified.

Field amplitude and phase are redefined under the scaling (6): $\Psi\leftrightarrow\tilde\Psi$, $\varphi\leftrightarrow\tilde\varphi+\alpha$. Christoffel symbols and Ricci tensor $R_{km}$ don't change. Vector $a_n$, metric tensor, mixed components of energy-momentum and Ricci tensors are multiplied by constant factors:
$$
a_n\leftrightarrow\tilde a_n,\ \ \ a^n\leftrightarrow\frac{U_0}{\tilde U_0}\tilde a^n,
$$
$$
g_{mn}\left(\psi\right)\leftrightarrow\frac{\tilde U_0}{U_0}\tilde g_{mn}\left(\tilde\psi\right),\eqno (17)
$$
$$
R_n^m\leftrightarrow\frac{U_0}{\tilde U_0}\tilde R_n^m,\ \ \ T_n^m\leftrightarrow\frac{U_0}{\tilde U_0}\tilde T_n^m.
$$
Therefore, Einstein's and Lagrange's equations don't change.

The Lagrangian (8) equals zero for the solution (12), whereas energy density is positive:
$$
E=\frac1{hc}\left(g^{mn}\frac{\partial\psi}{\partial x^m}\frac{\partial\psi^*}{\partial x^n}+U\left(\psi\psi^*\right)\right)=\frac2{hc}U_0\Psi^2>0.\eqno (18)
$$
The energy density (18) is constant, therefore the solution (12) corresponds to a stationary field condition.

Space-time volumes with the solution (12) type structure are similar to each other. Moreover, constant energy densities of these volumes may be related by the transformation (7). Therefore, they form a fractal set, the power law (2) is satisfied for them.

Mentioned above properties of the solution (12) including its stationarity and fractality are a consequence of the symmetry (5). Stationarity permits to refer this solution to the class of particle-like solution of the general relativity. Fractality implies that the solution corresponds to a set of noninteracting self-similar particles.

The phase path of the fields $\psi$ and $\psi^*$ is a circle (5):
$$
\psi\psi^*={\psi_1}^2+{\psi_2}^2=\Psi^2,
$$
$$
\psi=\psi_1+i\psi_2,\ \ \ \psi^*=\psi_1-i\psi_2.
$$
The function $\varphi$ is degree of rotation round the circle. Length of a circle arc i.e. interval of set $\left\{\psi_1,\psi_2\right\}$ equals
$$
dF^2=\left(d\psi_1\right)^2+\left(d\psi_2\right)^2=d\psi d\psi^*=\Psi^2\frac{\partial\varphi}{\partial x^m}\frac{\partial\varphi}{\partial x^n}dx^mdx^n.
$$
One can obtain a relation between the phase space interval $dF$ and the space-time interval $ds$:
$$
dF^2=\frac14\Psi^2\left[U_0ds^2-\left(a_m\frac{\partial\varphi}{\partial x^n}+a_n\frac{\partial\varphi}{\partial x^m}\right)dx^mdx^n\right].\eqno (19)
$$
The first equation (14) has the following solution:
$$
a_m=-3\frac{\partial\varphi}{\partial x^m}+d_m,
$$
where $d_m$ is a constant vector $\displaystyle \left(\frac{\partial d_m}{\partial x^n}=0\right)$. The expression (19) shows that the vector $d_m$ may be chosen so that the phase space interval is proportional to the time interval: $dF\sim dt$. Therefore, the solution (12) may describe a time-pulsating cosmological model. As energy density of the system is finite this model must be nonsingular. The example of such model is presented in the next section.
\vspace{1ex}

\begin{center}
{\bf 3. The cosmological model with meson field}
\end{center}

Let the fractal model is a model permitting existing of dependences of type (1 - 3). The model based on the solution (12) is an example of such model.

The solution (12) contains only the field derivatives $\displaystyle \frac{\partial\varphi}{\partial x^m}$, therefore it corresponds to both isotropic and anisotropic metrics. Below the homogeneous isotropic and conformally flat metric is used:
$$
ds^2=c^2dt^2-a^2\left[\left(dx^1\right)^2+\left(dx^2\right)^2+\left(dx^3\right)^2\right].
$$
Mixed Ricci tensor components and scalar curvature equal:
$$
R_0^0=-\frac1{c^2}\left[3\left(\frac{a_t}a\right)_t+3\left(\frac{a_t}a\right)^2\right],
$$
$$
R_1^1=R_2^2=R_3^3=-\frac1{c^2}\left[\left(\frac{a_t}a\right)_t+3\left(\frac{a_t}a\right)^2\right],
$$
$$
R=-\frac1{c^2}\left[6\left(\frac{a_t}a\right)_t+12\left(\frac{a_t}a\right)^2\right],
$$
where $a$ is scale factor of the model, index $t$ denotes partial derivative with cosmological time.

If we choose field Lagrangian in the form of
$$
L=\frac1{hc}\left(g^{mn}\frac{\partial\psi}{\partial x^m}\frac{\partial\psi^*}{\partial x^n}-U\left(\psi\psi^*\right)\right)+\frac{dF}{dt},
$$
where $\displaystyle \frac{dF}{dt}$ is a total derivative of some differentiable function, Einstein's equation (10) with the energy-momentum tensor
$$
T_{km}=\frac2{hc}\Psi^2\frac{\partial\varphi}{\partial x^k}\frac{\partial\varphi}{\partial x^m}-g_{km}\frac{dF}{dt}
$$
comes to the following two equations:
$$
3\left(\frac{a_t}a\right)^2=\frac{2\kappa}{hc}\Psi^2\left(\frac{\partial\varphi}{\partial t}\right)^2-\kappa c^2\frac{dF}{dt},
$$
$$
\frac1{c^2}\left[2\left(\frac{a_t}a\right)_t+3\left(\frac{a_t}a\right)^2\right]=-\frac{2\kappa}{hc}\Psi^2\frac1{a^2}\left(\frac{\partial\varphi}{\partial x}\right)^2-\kappa\frac{dF}{dt}.\eqno (20)
$$
Lagrange's equation (9) comes to two equations:
$$
\left(\frac1c\frac{\partial\varphi}{\partial t}\right)^2-\frac3{a^2}\left(\frac{\partial\varphi}{\partial x}\right)^2=U_0,
$$
$$
\frac1{c^2}\frac{\partial^2\varphi}{\partial t^2}-\frac3{a^2}\frac{\partial^2\varphi}{\partial x^2}+\frac3{c^2}\frac{a_t}a\frac{\partial\varphi}{\partial t}=0.\eqno (21)
$$
We have took into account here that $\displaystyle \frac{\partial\varphi}{\partial x^1}=\frac{\partial\varphi}{\partial x^2}=\frac{\partial\varphi}{\partial x^3}=\frac{\partial\varphi}{\partial x}$ in isotropic case. Four equations (20 - 21) determine four functions: $a$, $\displaystyle \frac{\partial\varphi}{\partial t}$, $\displaystyle \frac{\partial\varphi}{\partial x}$, $\displaystyle \frac{\partial^2\varphi}{\partial x\partial t}$. Equations (20) and the first equation (21) lead to the equation determining scale factor $a$:
$$
\left(\frac{a_t}a\right)_t+2\left(\frac{a_t}a\right)^2-\frac{\kappa c}{3h}U_0\Psi^2=-\frac{2\kappa c^2}{3}\frac{dF}{dt}.\eqno (22)
$$
The hyperbolic solution of this equation corresponds to the case of zero field Lagrangian and $\displaystyle \frac{dF}{dt}=0$:
$$
a=a_0\sqrt{\cosh{\left(\frac{t}{\tau}\right)}},
$$
where $\displaystyle \tau=1\left/\sqrt{\frac{2\kappa c}{3h}U_0\Psi^2}\right.$.

If we choose $\displaystyle \frac{dF}{dt}=\frac1{hc}U_0\Psi^2$ the equation has the periodic solution
$$
a=a_0\sqrt{\cos{\left(\frac{t}{\tau}+\phi\right)}}.\eqno (23)
$$
The period of function (23) equals $2\pi\tau$. The scale factor (23) turns into zero at the moment $t_*$ when $\displaystyle \frac{t_*}{\tau}+\phi=\frac{\pi}2\pm \pi n$. The solution (23) is not singular within the interval $\displaystyle 0\le \frac{t}{\tau}\le 2\pi$ if $\displaystyle \frac{t_*}{\tau}\ge 2\pi$. This condition permits to choose the phase: $\displaystyle \frac{\pi}2\pm \pi n-\phi >2\pi$. Therefore the model is not singular if
$$
\phi <-\frac{3\pi}2\pm \pi n.\eqno (24)
$$

The equation (22) has the integral for the periodic solution (23):
$$
\left(\frac{a_0}a\right)^4-4\tau^2\left(\frac{a_t}a\right)^2=1,
$$
which may be obtained through the expression
$$
\frac{a_t}a=-\frac1{2\tau}\tan{\left(\frac{t}{\tau}+\phi\right)}
$$
and the solution (23). The integral is a map of phase path (5) in the space-time.

Derivatives of the phase coordinate $\varphi$ equal:
$$
\frac1{c^2}\left(\frac{\partial\varphi}{\partial t}\right)^2=\frac14U_0\left[\left(\frac{a_0}a\right)^4+1\right],\ \ \ \frac1{a^2}\left(\frac{\partial\varphi}{\partial x}\right)^2=\frac1{12}U_0\left[\left(\frac{a_0}a\right)^4-3\right].
$$
One can define the metric tensor in the form analogous to the general definition (12) through these expressions. Further, the expressions imply that the parameter $a_0$ is a maximal scale factor value for the solution (23): $\displaystyle a\ge \frac1{\sqrt[3]{3}}a_0$. The comoving radial coordinate of the horizon equals:
$$
r\left(t\right)=\int\limits_0^t\frac{cdt}a=\frac{2c\tau}{a_0}F\left(\frac{\frac{t}{\tau}+\phi}2,2\right),
$$
where $\displaystyle F\left(\frac{\frac{t}{\tau}+\phi}2,2\right)$ is the elliptic integral of the first kind possessing recurring values with period $2\pi\tau$.

As the horizon comoving radial coordinate values are repeated, a model with pulsating space-time corresponds to the solution (23). This model is compacted, i. e. the total space volume is finite and the evolution in time is a periodic process of the space expansion and contraction. In the presence of the phase restriction (24) the space contracts to minimal nonzero volume. The two-dimensional analogy of such space-time is a torus with variable thickness where parallels are lines of time (lines of constant space coordinates) and meridians are space coordinate lines. Analogous compacted model has been constructed in the paper \cite{14} and possible astrophysical consequences of space volume finiteness are discussed there. It has been showed there that dynamical entropy of complex field is increasing during the space pulsating.
\vspace{3ex}

\begin{center}
{\bf 4. Conclusion}
\end{center}

The main results of this work are following. The cosmological model permitting physical explanation of the observational fractal properties of the galaxy and quasar distribution is constructed. Within the model framework these properties are consequences of the fractal properties of the initial density fluctuations spectrum and of the charged scalar meson matter field (complex field) rotary symmetry. This model is nonsingular; the Universe turns out to be compacted, pulsating and doubly-connected.

\end{document}